\documentclass[prl, twocolumn, showpacs, superscriptaddress]{revtex4-2}
\usepackage{amsmath, amssymb}
\usepackage{graphicx}
\renewcommand{\section}[1]{{\par\it #1.---}\ignorespaces}

\begin{document}
\title{Dissecting Superradiant Phase Transition in the Quantum Rabi Model}
\author{Yun-Tong Yang}
\affiliation{School of Physical Science and Technology, Lanzhou University, Lanzhou 730000, China}
\affiliation{Lanzhou Center for Theoretical Physics $\&$ Key Laboratory of Theoretical Physics of Gansu Province, Lanzhou University, Lanzhou 730000, China}
\author{Hong-Gang Luo}
\email{luohg@lzu.edu.cn}
\affiliation{School of Physical Science and Technology, Lanzhou University, Lanzhou 730000, China}
\affiliation{Lanzhou Center for Theoretical Physics $\&$ Key Laboratory of Theoretical Physics of Gansu Province, Lanzhou University, Lanzhou 730000, China}
\affiliation{Beijing Computational Science Research Center, Beijing 100084, China}

\begin{abstract}
The phase transition is both thermodynamically and quantum-mechanically ubiquitous in nature or laboratory and its understanding is still one of most active issues in modern physics and related disciplines. The Landau's theory provides a general framework to describe \textit{phenomenologically} the phase transition by the introduction of order parameters and the associated symmetry breakings; and is also taken as starting point to explore the critical phenomena in connection with phase transitions in renormalization group, which provides a complete theoretical description of the behavior close to the critical points. In this sense the microscopic mechanism of the phase transition remains still to be uncovered. Here we make a first attempt to explore the microscopic mechanism of the superradiant phase transition in the quantum Rabi model (QRM). We firstly perform a diagonalization in an operator space to obtain three fundamental patterns involved in the QRM and then analyze explicitly their energy evolutions with increasing coupling strengths. The characteristic behaviors found uncover the microscipic mechanism of the superradiant phase transition: one is active to drive the happening of phase transition, the second responses rapidly to the change of the active pattern and wakes up the third pattern to stablize the new phase. This kind of dissecting mechanism explains for the first time why and how happens the superradiant phase transition in the QRM and paves a way to explore the microscopic mechanism of the phase transitions happening popularly in nature.
\end{abstract}
\maketitle

\section{Introduction}
The finding of the critical phenomena can be dated back to 1822 due to de la Tour's observations of supercritical liquids \cite{Berche2009}, which have been further refined by Andrews in 1869 \cite{Rowlinson2003}, who defined the concept of the critical point. In 1873, Van der Waals proposed an equation of state, namely, the famous Van der Waals equation, to explain theoretically the continuity of the gaseous and liquid states of matter \cite{Waals1873, Rowlinson1973}. On the other hand, Pierre Curie found in 1895 that a ferromagnetic material tends to lose its permanent magnetism once the temperature increases up to certain value, named as Curie temperature or Curie point, and Weiss in 1907 explained it by his molecular mean-field theory \cite{Weiss1907}. In a binary alloy system, Bragg and Williams formulated the order-disorder transition in 1930's \cite{Bragg1934, Bragg1935}. In 1937, Landau presented a general and unified framework to treat \textit{phenomenologically} these phase transitions by introducing a concept of order parameter, associated with the breaking of certain symmetry \cite{Landau1937, Landau1980}, which formed the foundation of modern phase transition theory and of analyzing critical behaviors \cite{Chaikin2000, Sachdev2011}. While the concept of order parameter is still \textit{phenomenological} in nature underlying mean-field treatment, the scaling analysis and renormalization group method developed in 1960-70's further proposed the concept of the universality class of the phase transitions characterized by various critical exponents \cite{Wilson1974, Hahne1983}. Despite intensive and extensive studies of the details of various phase transitions in the past decades, in particular, along with the formulation of the renormalization group, it was clearly pointed out that ``\textit{Wilson’s theory for critical phenomena gave a complete theoretical description of the behaviour close to the critical point and gave also methods to calculate numerically the crucial quantities.}'' \cite{Nobel1982}. Based on the facts of the nature of phenomenology of the Landau's theory and the nature of description of the Wilson's theory, the microscopic mechanism of phase transitions still remains unexplored in a large extent \cite{Kastner2008}. Instead, the reason of phase transitions is usually attributed to the thermodynamical (at finite temperature) or quantum mechanical (at zero temperature) \textit{fluctuations} and the \textit{competitions} between various possible orders in thermodynamical \cite{Chaikin2000} or quantum phase transition \cite{Sachdev2011}. The questions about why and how happens the phase transition have not been answered in a sufficient and satisfactory way. 

In this paper we touch this question by taking the superradiant phase transition in the quantum Rabi model (QRM) \cite{Rabi1936, Rabi1937} as an example. An early analysis of the ground state in the QRM \cite{Ashhab2010, Hwang2010} showed that the ground state exhibits a squeezing in the strong, ultrastrong, even deep-strong coupling regime\cite{Wallraff2004, Niemczyk2010, Casanova2010, Yoshihara2017, Yoshihara2018, Mueller2020}, a precursor of distinguished physics of superradiance, which has been further confirmed theoretically \cite{De2013, Hwang2015, Ying2015, Liu2017, Wang2018Y, Peng2019, Zhu2020, Garbe2020, Jiang2021, Zhuang2021, Stransky2021, Yang2022}. Very recently, this phase transition has been observed in a single trapped ion \cite{Cai2021, Cai2022CPL} and stimulated experimentally in the platform of nuclear magnetic resonance \cite{Chen2021}. In Ref. \cite{Yang2022}, the present authors pointed out that the superradiant phase transition happens due to the induced double-well potenial in a full quantum mechanical way as increasing coupling strength across its critical point and here we further explore why and how the phase transition happens. 

Methodologically, different to that in Ref. \cite{Yang2022}, here we firstly make a diagonalization in an operator space (see below) to obtain three fundamental patterns involved in the QRM, which are exact in a sense that the Hamiltonian matrix can be exactly reproduced for a given Fock basis. In order to further confirm the validity of the pattern picture, we compare the result we present with that obtained by numerical exact diagonalization. Then, we analyze the ground-state superradiant phase transition by checking the energy levels of the ground state and the first excited state, which become almost degenerate in the superradiant phase. It is found that these three fundamental patterns play quite different roles in the process of the superradiant phase transition: one is active to drive the happening of the phase transition by lowering rapidly energy of the system as the coupling strength approaches to its critical point. The second one is found to response rapidly to the change of the active pattern and to wake up the third pattern in order to fix the change of energy. And finally, the wakened third pattern begins to balance the first one. Thus the third pattern is found to be passive, which plays a role to stabilize the new phase as continually increasing the coupling strengths. As a consequence, the second pattern plays a role of inspector. It always does accompany with, thus is sensitive to, the change of the system. The different roles played by these three patterns uncover why and how happens the superradiant phase transition in the QRM, which is our meaning that dissects the superradiant phase transition. Both methodologically and physically, our result paves a way to explore the phase transitions happening in other systems. 

\section{Model and Method}
\begin{figure}[tbp]
\begin{center}
\includegraphics[width = \columnwidth]{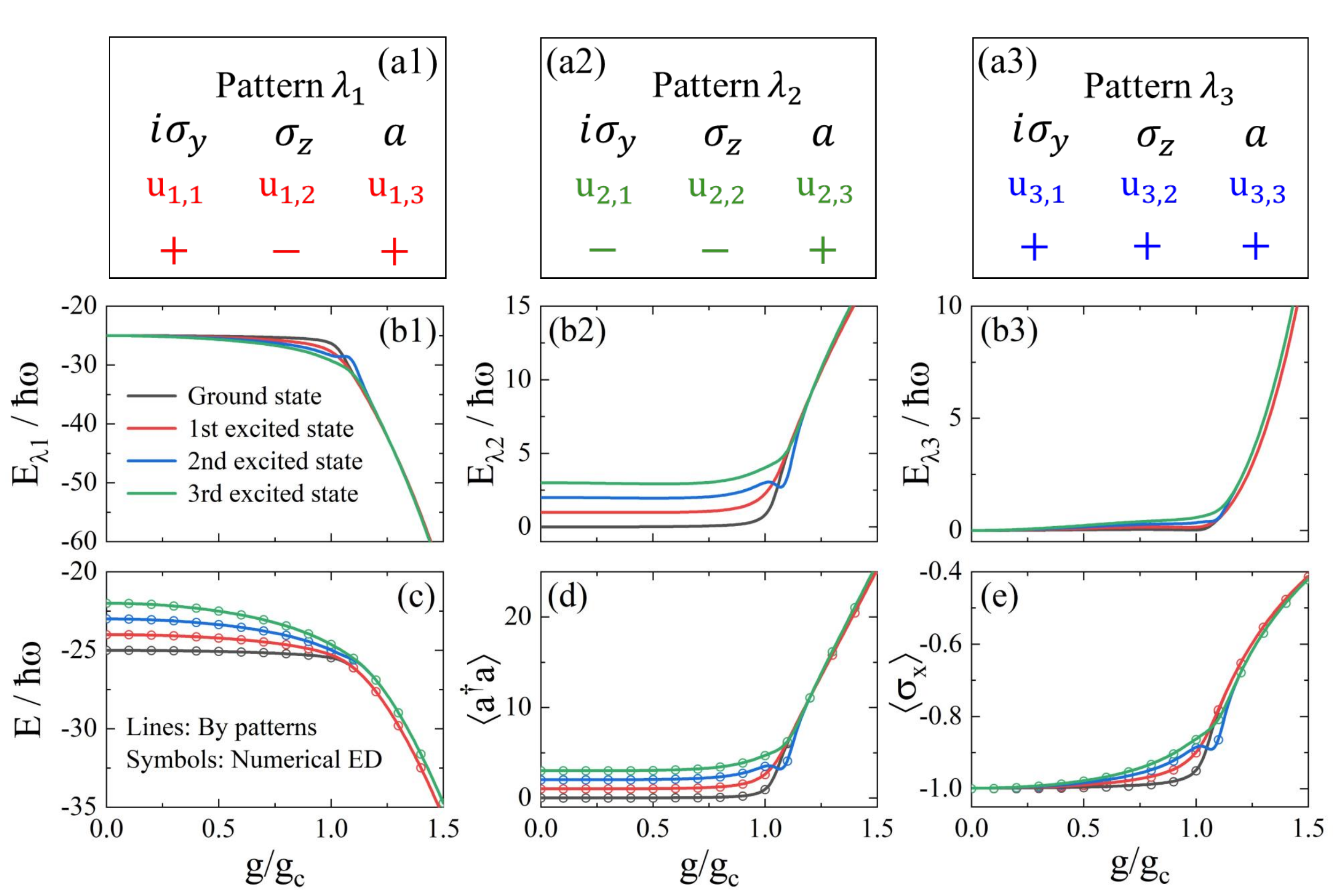}
\caption{The marks of the obtained patterns and comparisons of the physical quantities with the results obtained by numerical exact diagnoalization. (a1)-(a3) Patterns marked by $\lambda_1, \lambda_2$ and $\lambda_3$ and their wavefunctions with relative phase denoted by $\pm$. Note that there is a total free phase factor $e^{i\pi}$, which is omitted here for simplicity. (b1)-(b3) The first four pattern energy levels as functions of the coupling strengths rescaled by $g_c = \sqrt{1+\sqrt{1+\frac{\Delta^2}{16}}}$ \cite{Ying2015} and $\Delta = 50$ is taken here and hereafter. (c) The summations of the pattern energy levels (lines) and their comparison with the results obtained directly by numerical exact diagonalization (symbols). (d) Comparison of the summations of the pattern's photon numbers $\langle a^\dagger a \rangle_{\lambda_n}(n=1,2,3)$ (not shown) to those obtained by numerical exact diagonalization (symbols). (e) Comparison of the summations of the pattern's $\langle \sigma^x\rangle_{\lambda_n}(n=1,2,3)$(not shown) to those obtained by numerical exact diagonalization (symbols). The comparisons are made for the first four energy levels.}\label{fig1}
\end{center}
\end{figure}

The Hamiltonian of the QRM is standard, it reads  
\begin{equation}
\hat{H} = \hbar \omega \left(\hat{a}^\dagger \hat{a} + \frac{\Delta}{2} \hat{\sigma}_x +  g\left(\hat{a} + \hat{a}^\dagger\right) \hat{\sigma}_z\right), \label{Rabi0}
\end{equation}
where $\hat{a}^\dagger (\hat{a})$ is creation (destruction) operator of the single photon mode and $\hat{\sigma}_x, \hat{\sigma}_z$ are usually Pauli matrices denoting the two-level atom. For convenience, we rescale the Hamiltonian by the mode frequency $\hbar\omega$, thus the energy interval $\Delta$ and the coupling strength $g$ used in the following are dimensionless, and below the units of energy $\hbar\omega$ is omitted for simplicity. By using the relation $\hat{\sigma}_y \hat{\sigma}_z = i\hat{\sigma}_x$, Eq. (\ref{Rabi0}) can be reformulated as follows
\begin{eqnarray}
\hat{H} &=&
\left(
\begin{array}{ccc}
-i\hat{\sigma}_y & \hat{\sigma}_z &\hat{a}^\dagger 
\end{array}
\right)
\left(
\begin{array}{ccc}
 0 & \frac{\Delta}{4} & 0 \\
\frac{\Delta}{4} & 0 & g \\
 0 & g & 1
\end{array}
\right)
\left(
\begin{array}{c}
 i\hat{\sigma}_y \\
 \hat{\sigma}_z \\
  \hat{a} 
\end{array}
\right) \label{Rabi1a}\\
&=& \sum_{n=1}^3 \lambda_n \hat{A}^\dagger_n \hat{A}_n, \label{Rabi1b}\\
\hat{A}_n &=& u_{n,1}\left(i\hat{\sigma}_y\right) + u_{n,2}\hat{\sigma}_z + u_{n,3}\hat{a},\label{Rabi1c}
\end{eqnarray}
where $\{\lambda_n,u_{n}\}(n = 1,2,3)$ are eigenvalues and eigenfunctions of the matrix in Eq. (\ref{Rabi1a}), forming three basic patterns marked by the eigenenergy $\lambda_n$ and their wavefunction components, as shown in Fig. \ref{fig1} (a1)-(a3). Before discussing the explicit physics of the phase transition, it is necessary to verify the validity of the pattern picture obtained. 

\section{Patterns and Solution} 
Eq. (\ref{Rabi1b}) can be solved by inserting into the complete basis $|\sigma_z,m\rangle$, where $\hat{\sigma}_z|\sigma_z\rangle = \pm(\uparrow,\downarrow)|\sigma_z\rangle$ denoting the spin eigenstate along $z$-direction and $\hat{a}^\dagger \hat{a}|m\rangle = m|m\rangle (m=0,1,\cdots,N)$ denoting the truncated Fock basis with photon number $m$. Firstly, one obtains the matrix $\left[\hat{A}_n\right]_{\sigma_z,m;\sigma_{z}^{\prime},m^{\prime}} = \langle\sigma_z, m|\hat{A}_n|m^{\prime},\sigma_{z}^{\prime}\rangle$ and then Eq. (\ref{Rabi1b}) can be solved by diagonalizing the matrix with matrix elements
\begin{equation}
\langle \sigma_z, m|\hat{H}|m^{\prime},\sigma_{z}^{\prime}\rangle = \sum_{n=1}^3 \lambda_n \sum_{\sigma_{z}^{\prime\prime},q} \left[\hat{A}^\dagger_n\right]_{m,\sigma_z;q,\sigma_{z}^{\prime\prime}}\left[\hat{A}_n\right]_{q,\sigma_{z}^{\prime\prime};m^{\prime},\sigma_{z}^{\prime}}.\label{Rabi2}
\end{equation}
Figure \ref{fig1} (b1)-(b3) present the first four energy levels for all three patterns by taking truncated Fock basis with $N = 200$, which is sufficient to discuss the physics here we are interested in. It is noticed that a dramatic energy change occurs in all three patterns around the coupling strength $g\sim g_c$, which means the happening of superradiant phase transition in the system, a well-known result in the literature \cite{Hwang2015, Liu2017}. The validity of our formulation is confirmed by comparing the summation (solid lines) of respective physical quantities for all three patterns, for example, the energies shown in (b1)-(b3), to those obtained directly by numerical exact diagonalization (symbols). The results are exactly same, as shown in Fig.\ref{fig1} (c) for the energy levels, (d) for the photon number $\langle \hat{a}^\dagger \hat{a}\rangle$, as well as (e) for the spin-flip $\langle \hat{\sigma}_x\rangle$ (for the latter two quantities the respective pattern components have not been shown here). It is not surprising since no any additional approximation has been introduced in our pattern formulation in comparison to the numerical exact diagonalization besides truncated Fock basis. Thus our pattern formulation gives an alternative way to dissect Hamiltonian of the QRM into fundamental patterns, which provide a novel angle to analyze the physics involved in the QRM, for example, the superradiant phase transition we focus on here. 

\begin{figure}[tbp]
\begin{center}
\includegraphics[width = \columnwidth]{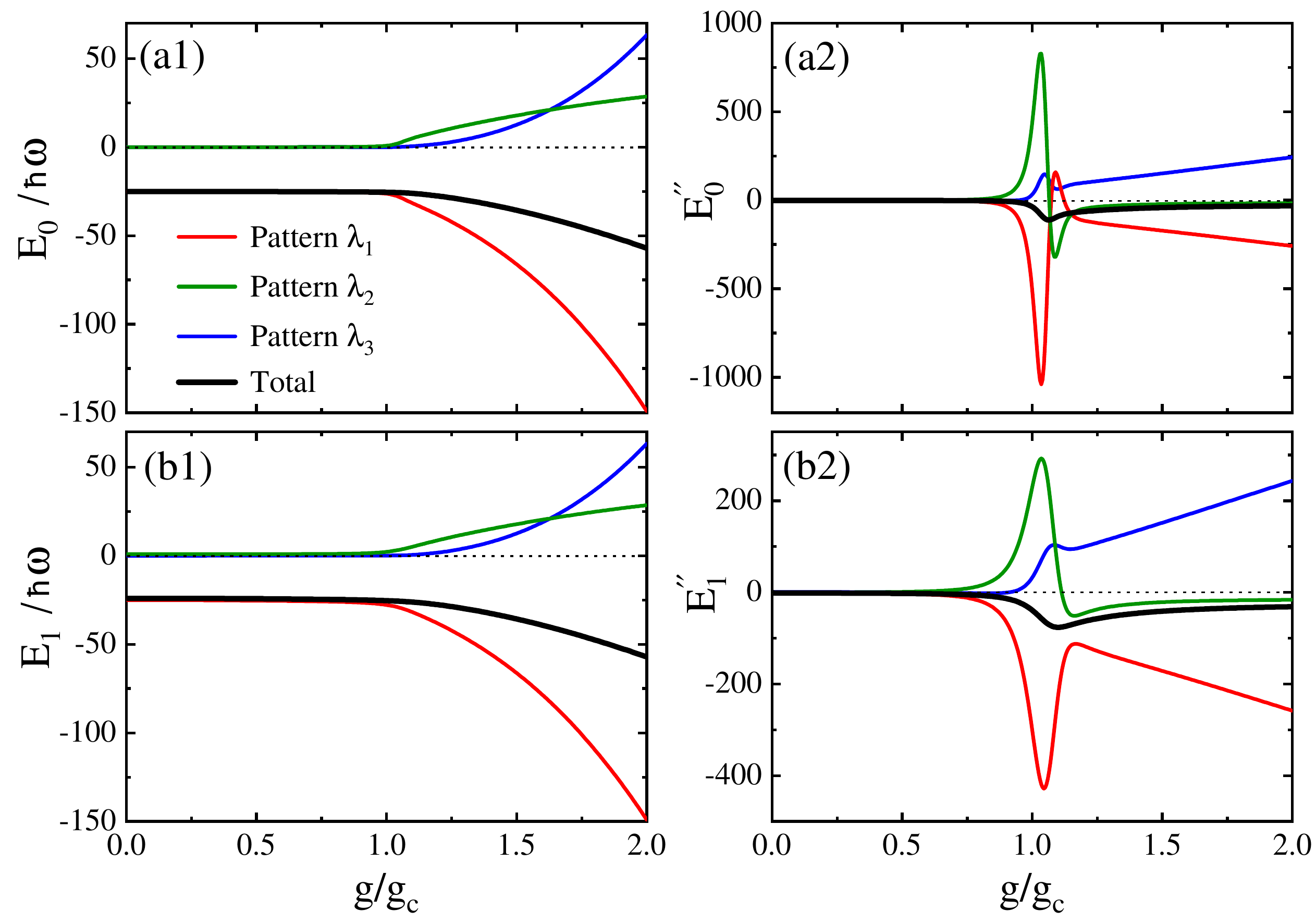}
\caption{(a1) $\&$ (b1) The energy levels of the ground state and the first excited state (heavy black solid ines) and corresponding pattern components (thin red, green, and blue solid lines) as functions of the coupling strength. (a2) $\&$ (b2) The second-order derivatives of the corresponding energy levels (heavy black solid lines) and their pattern components (thin red, green, and blue solid lines). }\label{fig2}
\end{center}
\end{figure}

\section{The nature of phase transition} 
It is well-known, as also mentioned above, that the QRM involves a superradiant phase transition as the coupling strength $g/g_c \sim 1$, where the energy levels, the photon numbers as well as the spin-flip have a dramatic change, as shown in Fig. \ref{fig1}(c)-(e), respectively. Here we take the ground-state and first-excited state energy levels to make a detailed and deeper analysis why and how happens the phase transition by checking carefully the behaviors of the patterns. As shown in Fig. \ref{fig2} (a1), the ground state energy components for three patterns and, as a consequence, the ground state energy keep almost unchanged in the normal phase for the weak coupling regime. An apparent downturn happens firstly and then descends continually for the pattern $\lambda_1$ (thin red line). At the same time, the second pattern $\lambda_2$ makes a rapid response to the change of the pattern $\lambda_1$ in a compensation way, as a result, the total energy of the system (heavy black line) remains almost unchanged in a short interval. With further increasing the coupling strengths, the third pattern $\lambda_3$ begins to response in a slightly slow but more smooth way, but its weight increases more rapidly and goes beyond that of the pattern $\lambda_2$ quickly and finally is able to balance roughly the energy change of the pattern $\lambda_1$. As a result, the ground state energy level begins to descend in a more smooth way, as shown in heavy black solid line in Fig. \ref{fig2}(a1). In order to see more clearly what happens in such a process, we check the second-order derivatives of all these energies with respect to the coupling strength. The results are shown in Fig.\ref{fig2} (a2), from which one check how happens the phase transition. Obviously, with increasing the coupling strength, the change is launched firstly by the pattern $\lambda_1$ by requiring energy rapidly, a prompt response is followed by the pattern $\lambda_2$ in a way against to this change. At the same time, it also wakens up the pattern $\lambda_3$ to take part in the balance of the system. Once the pattern $\lambda_3$ is involved, the trends of energy changes for the patterns $\lambda_1$ and $\lambda_2$ are changed, the pattern $\lambda_1$ turns back and the pattern $\lambda_2$ is relaxed. With further increasing the coupling strength, the pattern $\lambda_3$ plays a major role to balance the pattern $\lambda_1$. As a result, a newly developed phase is stabilized and in this process the total energy changes in a much smooth way, as shown in Fig.\ref{fig2} (a2) as the heavy black line. More interestingly, the pattern $\lambda_2$ goes along with the total state of the system, as shown in Fig.\ref{fig2} (a2) as thin green solid lines. 

From the above analysis we can summarize the roles played by these three patterns in the process of phase transition: the pattern $\lambda_1$ is active, playing a role of initiator of the phase transition, and the pattern $\lambda_3$ is passive, playing a role of stabilizer of the newly developed phase, namely, the superradiant phase; more interestingly, the pattern $\lambda_2$ is very sensitive to the state change of the system, and it plays a role of inspector, who takes responsibility to probe and response promptly to the state change of the system, and to waken up the pattern $\lambda_3$ to begin to work in order to stabilize the newly developed phase. A completely similar but slightly smooth process also happens in the first excited state, as shown in Fig.\ref{fig2} (b2). This result explains why and how happens the well-known superradiant phase transition, which is main result of the present work. 

\begin{figure}[tbp]
\begin{center}
\includegraphics[width = \columnwidth]{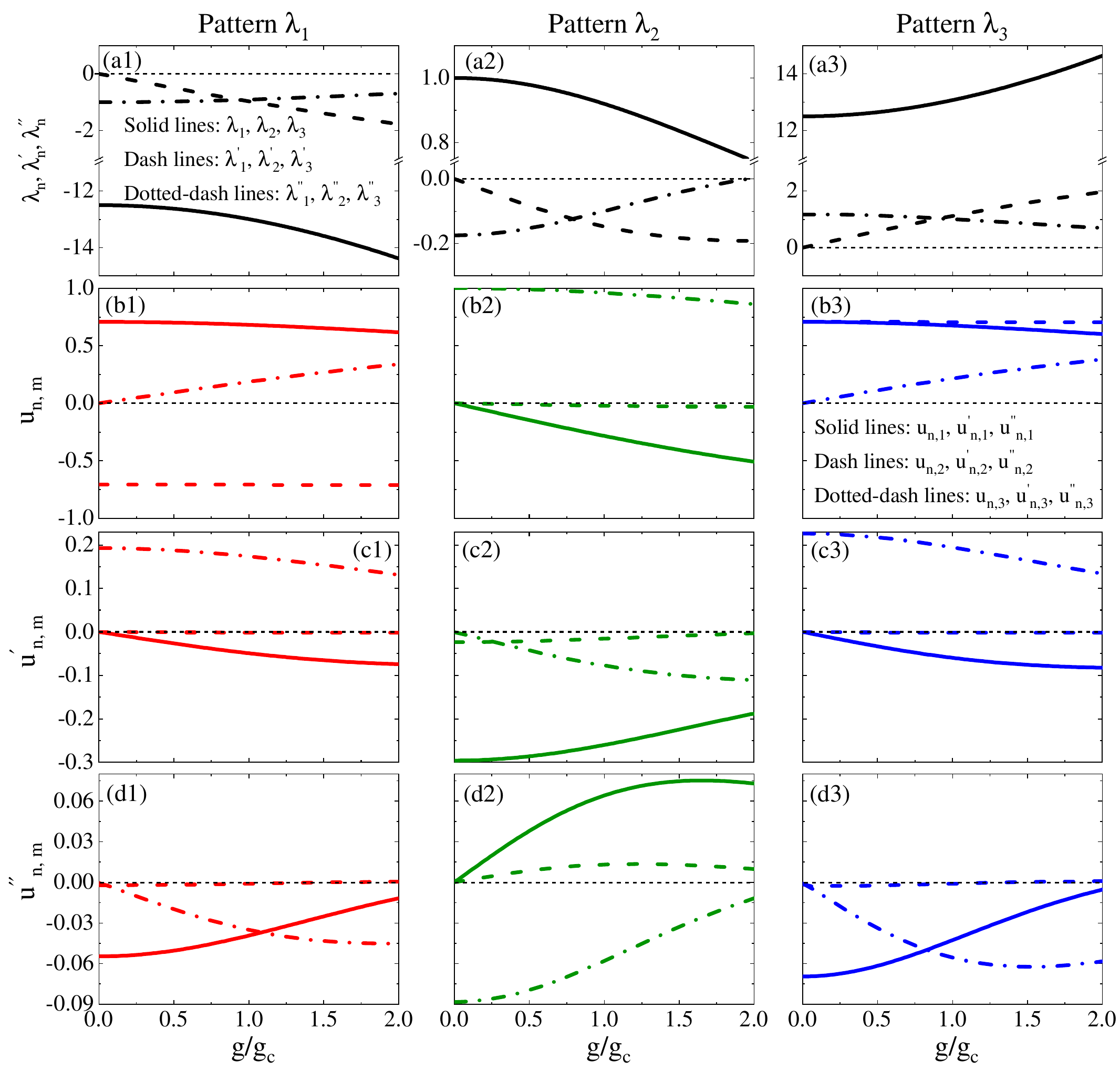}
\caption{Patterns marked by their eigenenergies and eigenfunctions as functions of the coupling strengths. The first column: (a1) the eigenenergy of the pattern $\lambda_1$ and its first- and second-order derivatives with respect to the coupling strength; (b1) the eigenfunctions $u_{1,m}$($m=1,2,3$), (c1) their first-order derivative, and (d1) their second-order derivative for the pattern $\lambda_1$ (red). The second column: (a2) the eigenenergy of the pattern $\lambda_2$ and its first- and second-order derivatives with respect to the coupling strength; (b2) the eigenfunctions $u_{2,m}$($m=1,2,3$), (c2) their first-order derivative, and (d2) their second-order derivative for the pattern $\lambda_2$ (green). The third column: (a3) the eigenenergy of the pattern $\lambda_3$ and its first- and second-order derivatives with respect to the coupling strength; (b3) the eigenfunctions $u_{3,m}$($m=1,2,3$), (c3) their first-order derivative, and (d3) their second-order derivative for the pattern $\lambda_3$ (blue). For each pattern, there are three components corresponding to the operators $i\hat{\sigma}_y$(solid lines), $\hat{\sigma}_z$(dash lines) and $\hat{a}$(dotted-dash lines).}\label{fig3}
\end{center}
\end{figure}

After discussing the roles played by these three patterns in the process of the phase transition, it is useful to further explore the properties of the patterns. In Fig. \ref{fig3} we present eigenenergies and corresponding eigenfunctions of the patterns and their first- and second-order derivatives as functions of the coupling strengths in order to further interpret their roles played in the process of the phase transition. A few properties are noted: (i) the signs or relative phases between the components in each pattern remain fixed, which is sufficient and necessary conditions to mark the patterns; (ii) in the patterns $\lambda_1$ and $\lambda_3$, the two-level dominates, but the weight of the photon mode increases continually with the increasing coupling strengths. In addition, in the patterns $\lambda_1$ and $\lambda_3$, the weight of the component $i\hat{\sigma}_y$ decreases slowly but continually with increasing the coupling strength, but that of the component $\hat{\sigma}_z$ has no significant change, which reflects the fact that the patterns $\lambda_1$ and $\lambda_3$ define the states of the system;  (iii) in the pattern $\lambda_2$, the photon mode is dominant and the weight of the first component $i\hat{\sigma}_y$ increases continually with increasing the coupling strength, which is understandable since the spin-flip is associated with the absorption or emission of the photons, and that of the second component $\hat{\sigma}_z$ remains negligible in the whole coupling regime we considered, which reflects the inspector role played by the pattern $\lambda_2$. The above analysis can be further confirmed by checking the first and second derivatives of the pattern eigenfunctions as function of the coupling strengths, as shown in Fig. \ref{fig3} (c1)-(c3) and (d1)-(d3). In particular, the component $\hat{\sigma}_z$ defines the states of the two-level in the pattern $\lambda_1$ and $\lambda_3$, which remains almost unaffected as changing the coupling strengths. In contrast, the component $i\hat{\sigma}_y$ and the photon mode are active as the coupling strength is changed, which is physically reasonable. In this sense the phase transition is a dynamical process in nature. 

\begin{figure}[tbp]
\begin{center}
\includegraphics[width = \columnwidth]{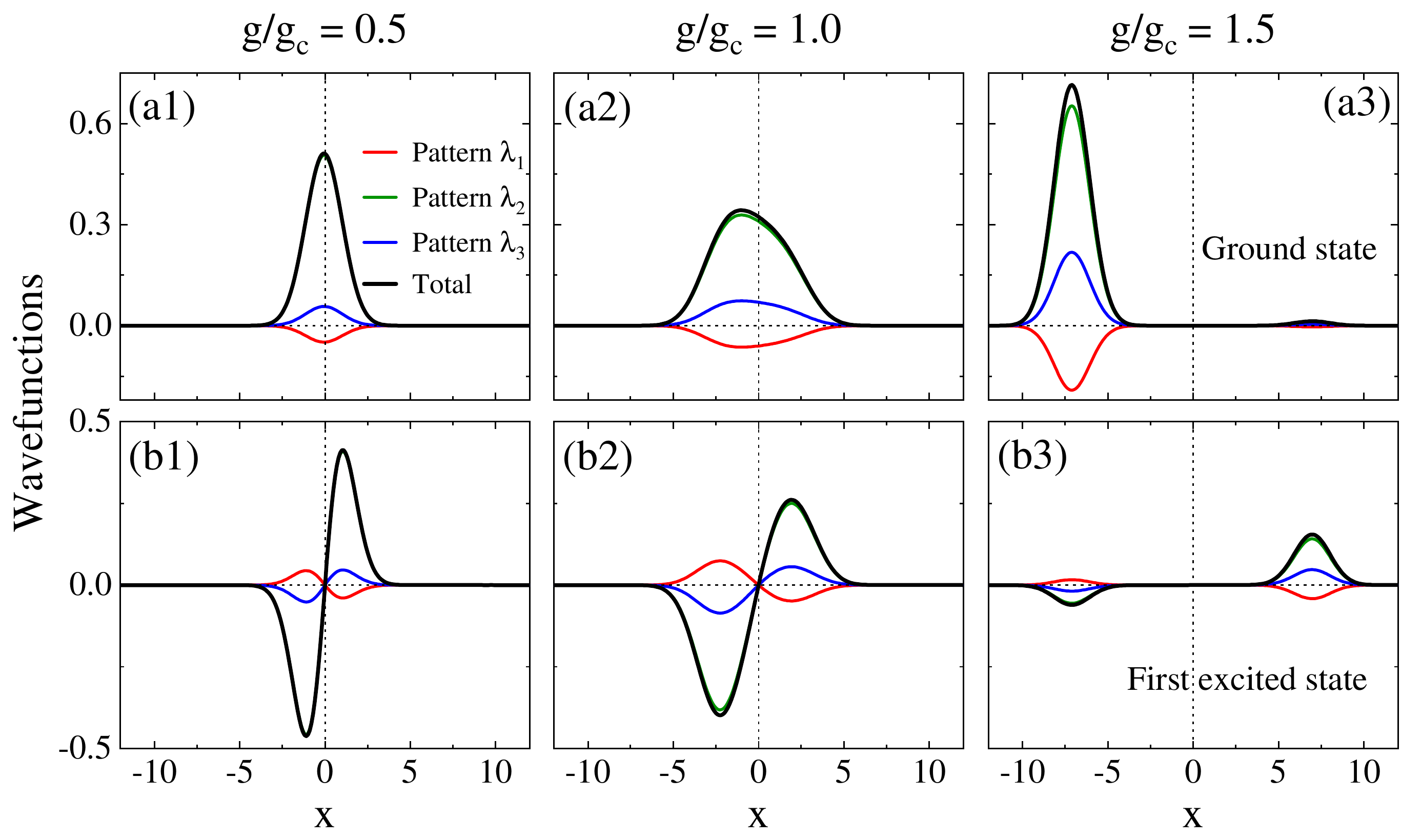}
\caption{The wavefunctions of the ground state (a1)-(a3) and the first excited state (b1)-(b3) for three typical coupling strengths $g/g_c = 0.5$ (a1) $\&$ (b1); $1.0$ (a2) $\&$ (b2); $1.5$ (a3) $\&$ (b3), corresponding to the normal, critical and superradiant phase, respectively, in the Fock basis. The heavy black solid lines denote the total wavefunctions and the thin colored lines (red, green and blue) correspond to the three components in patterns $\lambda_1$, $\lambda_2$ and $\lambda_3$, respectively, }\label{fig4}
\end{center}
\end{figure}

\begin{figure}[h]
\begin{center}
\includegraphics[width = \columnwidth]{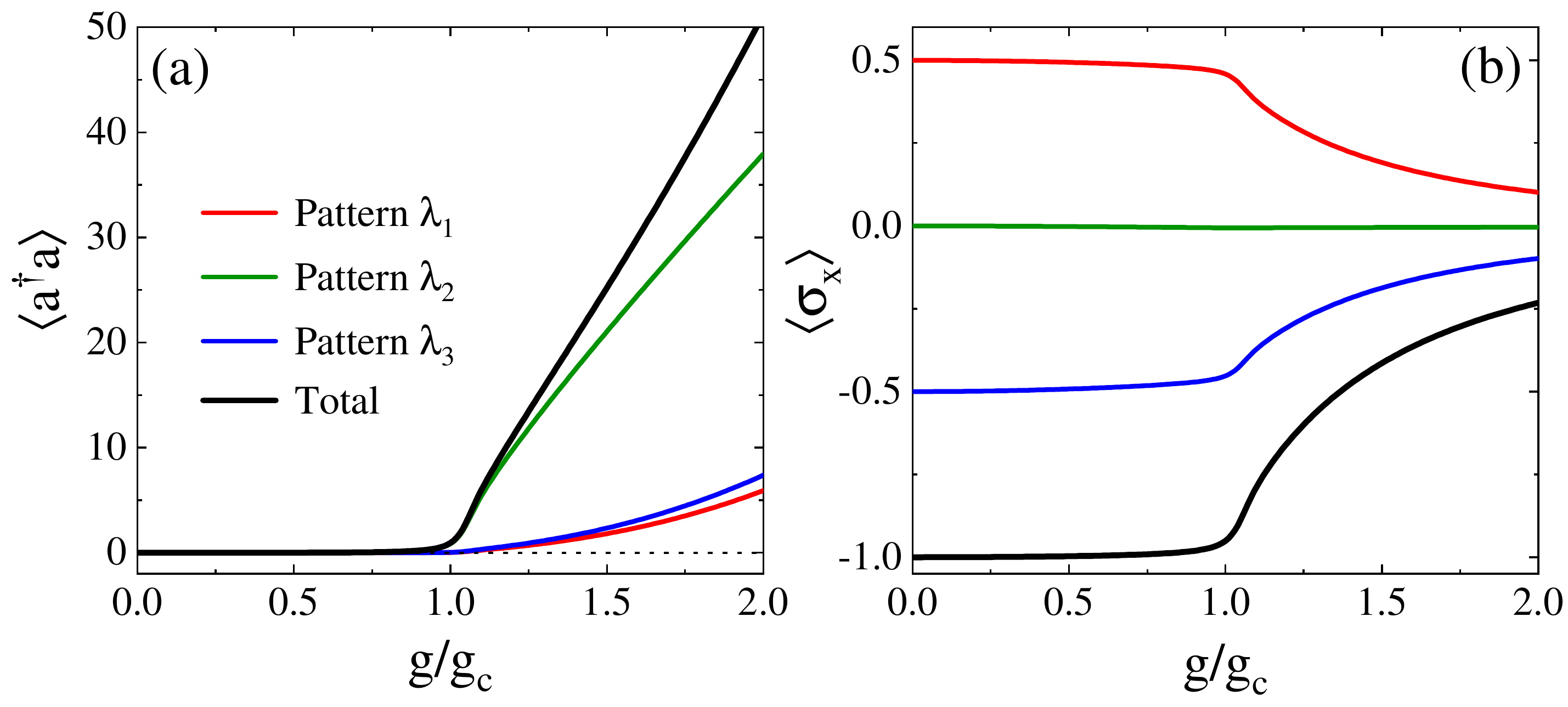}
\caption{The total photon number (a) and the spin-flip (b), and the corresponding components in patterns for the ground state as functions of the coupling strength. The black solid lines denote total one and the red, green, and blue thin solid lines correspond to the components in patterns $\lambda_1$, $\lambda_2$, and $\lambda_3$, respectively.}\label{fig5}
\end{center}
\end{figure}

It is also interesting to see the wavefunctions of the system, consisting of the direct product of the two-levels or the up- and down-spins and the Fock basis of the photon mode. Here we only show the up-spin branch and the goal is to show how evolve the photon mode components in the three patterns in comparsion to the total wavefunctions. Fig. \ref{fig4} shows the results for the ground state (the first row) and the first excited state (the second row). In the normal phase, it is noticed that the pattern $\lambda_2$ dominates in a way that is almost the same as the total one and the other two patterns are excited slightly but in the opposite way, thus have less contribution to the total wavefunction. The situation continues up to the critical point around $g/g_c \sim 1$ and the total wavefunction and its pattern components begin to become more asymmetric. Entering into the superradiant phase at the typical strong coupling regime, the total wavefunction and their pattern components are very asymmetric for both the ground state and the first excited state. Moreover, the weights of the patterns $\lambda_1$ and $\lambda_3$ increase but still remain out of phase, as a result, the pattern $\lambda_2$ still dominates over the other two patterns. In addition, two interesting features are observed, one is that the pattern $\lambda_1$ is always in the opposite to the pattern $\lambda_3$, and the second is that the pattern $\lambda_3$ is always in the same phase as the pattern $\lambda_2$, a dominant carrier of the superradiant photons, as shown in Fig. \ref{fig5} (a). In a word, the patterns $\lambda_1$ and $\lambda_3$ have a compentition relationship, but the pattern $\lambda_3$ and $\lambda_2$ have a cooperative relationship dominated by the in-phase in the two-level, as shown in Fig. \ref{fig1} (a2) and (a3). This picture is in consistent with the roles played by the patterns, as uncovered in Fig. \ref{fig2}. In line with that the pattern $\lambda_1$ and $\lambda_3$ carry few photons, the pattern $\lambda_2$ has a negligible contribution to the spin-flip, as shown in Fig. \ref{fig5} (b). These results clearly indicate the fact that with increasing the coupling strength the pattern $\lambda_1$ drives the happening of the superradiant phase transition, the pattern $\lambda_2$ carries most of the superradiant photons, and the pattern $\lambda_3$ stabilizes the superradiant phase in against to the pattern $\lambda_1$. The role dissection of the patterns in the dynamical process of the phase transitions provides a novel insight on the understanding the phase transition and the related physics.

\section{Summary}
Through the diagonalization in operator space we obtain a pattern formulation in order to dissect the process of the superradiant phase transition in the QRM. Three patterns and their roles played in the superradiant phase transition are clearly uncovered: one drives the happening of the phase transition, and the second is sensitive to the phase transition and carrys the superradiant photons, and the third stabilizes the newly developed superradiant phase in against to the phase transition driver. Our result answers for the first time why and how happens the superradiant phase transition. Due to the fundamental role of the QRM in describing the interaction between light and matter, it is expectable that the dissection of the patterns presented in the present work can be further extended to more complex models such as Dicke \cite{Dicke1954} and spin-boson model \cite{Leggett1987} in order to understand more intriguing physics involved in these models such as decoherence and dissipation mechanisms. 

\section{Acknowledgments}
The work is partly supported by the National Key Research and Development Program of China (Grant No. 2022YFA1402704) and the programs for NSFC of China (Grant No. 11834005, Grant No. 12047501).



%

\end{document}